\documentclass[twocolumn,letterpaper,amsmath,amssymb,floatfix,aps,superscriptaddress]{revtex4}

\usepackage{graphicx}
\usepackage{dcolumn}
\usepackage{bm}
\usepackage[usenames]{color}
\usepackage{hyperref}
\usepackage{ulem}

\begin{document}

\title{Theory of polymer translocation through a flickering nanopore under an alternating driving force}

\author{Jalal Sarabadani}
\email{jalal.sarabadani@aalto.fi}
\affiliation{Department of Applied Physics and COMP Center of Excellence, Aalto University School of Science, 
P.O. Box 11000, FI-00076 Aalto, Espoo, Finland}

\author{Timo Ikonen}
\affiliation{~VTT Technical Research Centre of Finland Ltd., P.O. Box 1000, FI-02044 VTT, Finland}

\author{Tapio Ala-Nissila}
\affiliation{Department of Applied Physics and COMP Center of Excellence, Aalto University School of Science, 
P.O. Box 11000, FI-00076 Aalto, Espoo, Finland}
\affiliation{Department of Physics, Box 1843, Brown University, Providence, Rhode Island 02912-1843.}

\begin{abstract}
We develop a theory for polymer translocation driven by a time-dependent force 
through an oscillating nanopore. To this end, we extend the iso-flux tension propagation theory (IFTP) 
[Sarabadani \textit{et al., J. Chem. Phys.}, 2014, \textbf{141}, 214907] for such a setup. We assume 
that the external driving force in the pore has a component oscillating in time, and the flickering pore 
is similarly described by an oscillating term in the pore friction. In addition to numerically solving 
the model, we derive analytical approximations that are in good agreement with the numerical simulations. 
Our results show that by controlling either the force or pore oscillations, the translocation process can 
be either sped up or slowed down depending on the frequency of the oscillations and the characteristic 
time scale of the process. We also show that while in the low and high frequency limits the translocation 
time $\tau$ follows the established scaling relation with respect to chain length $N_0$, in the intermediate 
frequency regime small periodic fluctuations can have drastic effects on the dynamical scaling. The results 
can be easily generalized for non-periodic oscillations and elucidate the role of time dependent forces and 
pore oscillations in driven polymer translocation.
\end{abstract}

\maketitle

\section{Introduction} \label{introduction}

\begin{figure*}[t]\begin{center}
    \begin{minipage}[b]{0.31\textwidth}
        \includegraphics[width=1.0\textwidth]{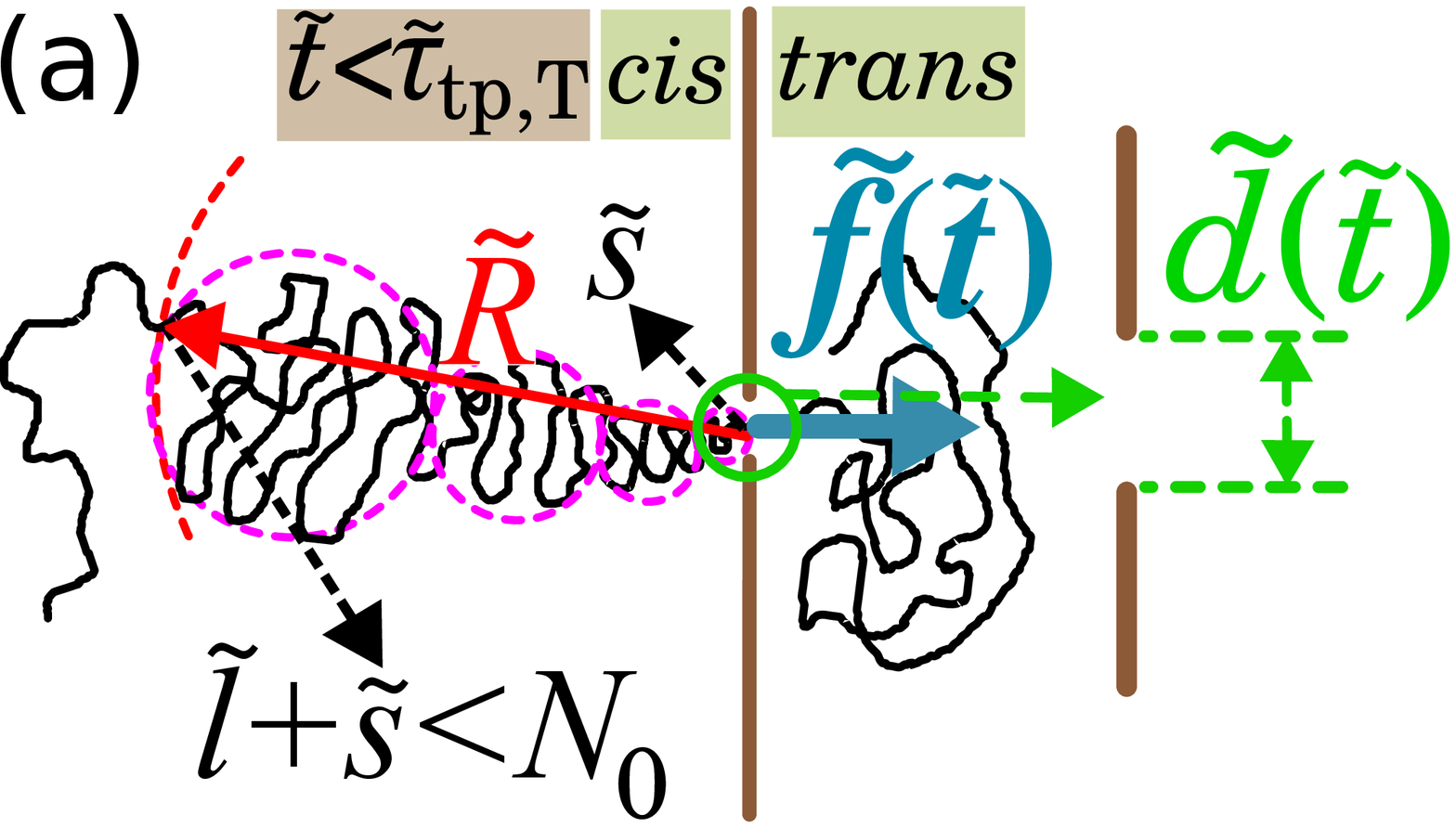}
    \end{minipage} \hskip+0cm
    \begin{minipage}[b]{0.22\textwidth}
        \includegraphics[width=1.0\textwidth]{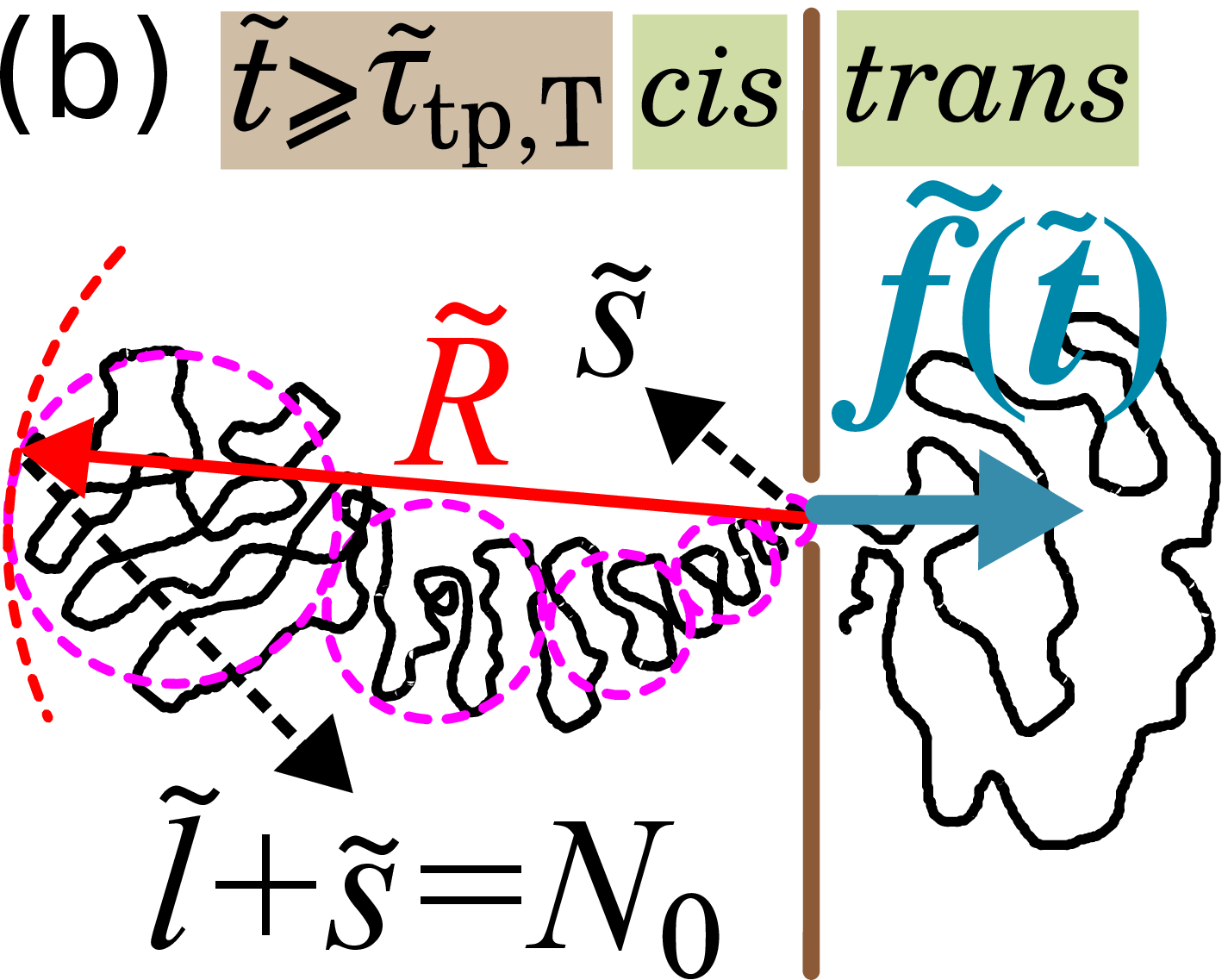}
    \end{minipage} \hskip0cm
    \begin{minipage}[b]{0.22\textwidth}
        \includegraphics[width=1.0\textwidth]{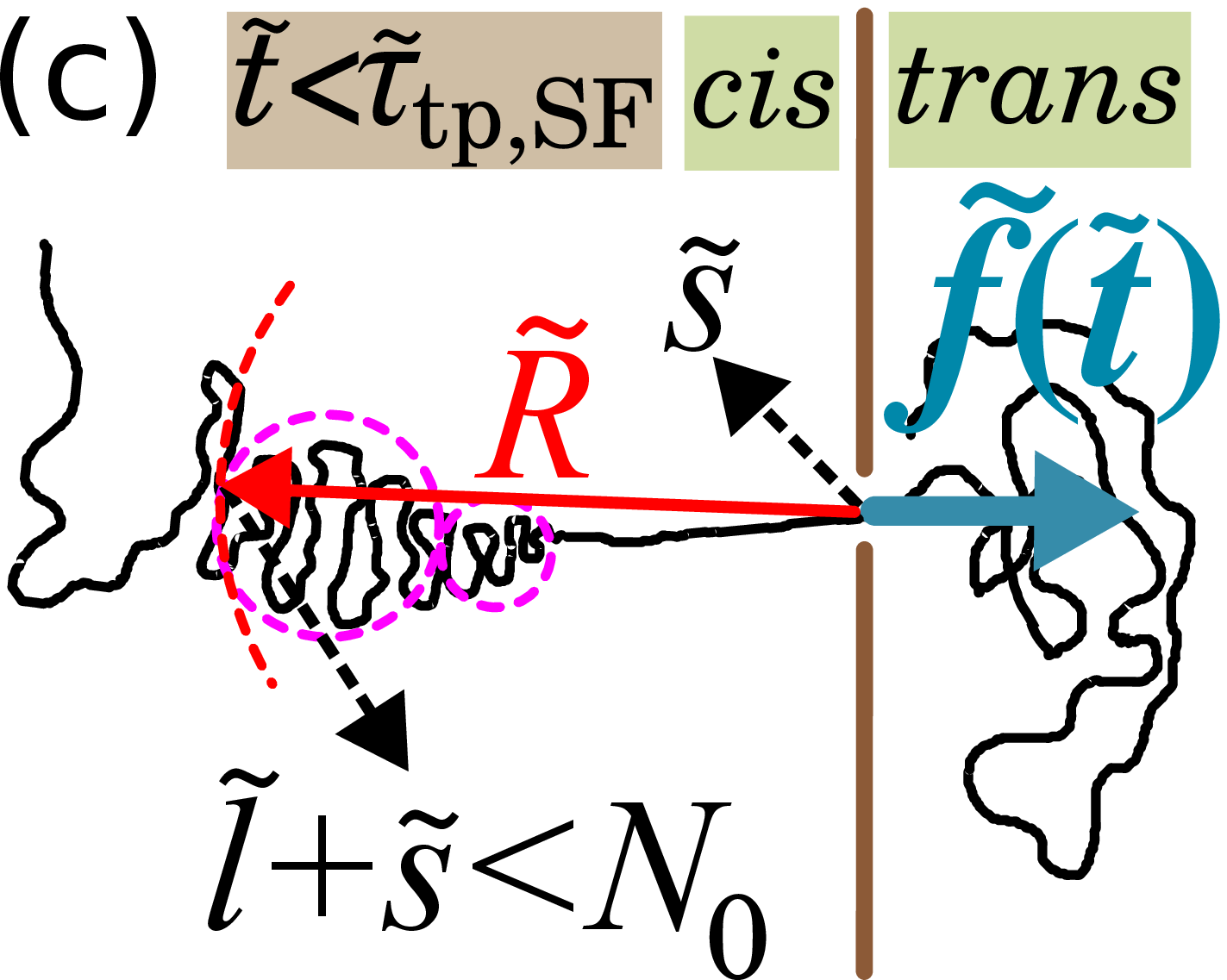}
    \end{minipage} \hskip0cm
    \begin{minipage}[b]{0.22\textwidth}
        \includegraphics[width=1.0\textwidth]{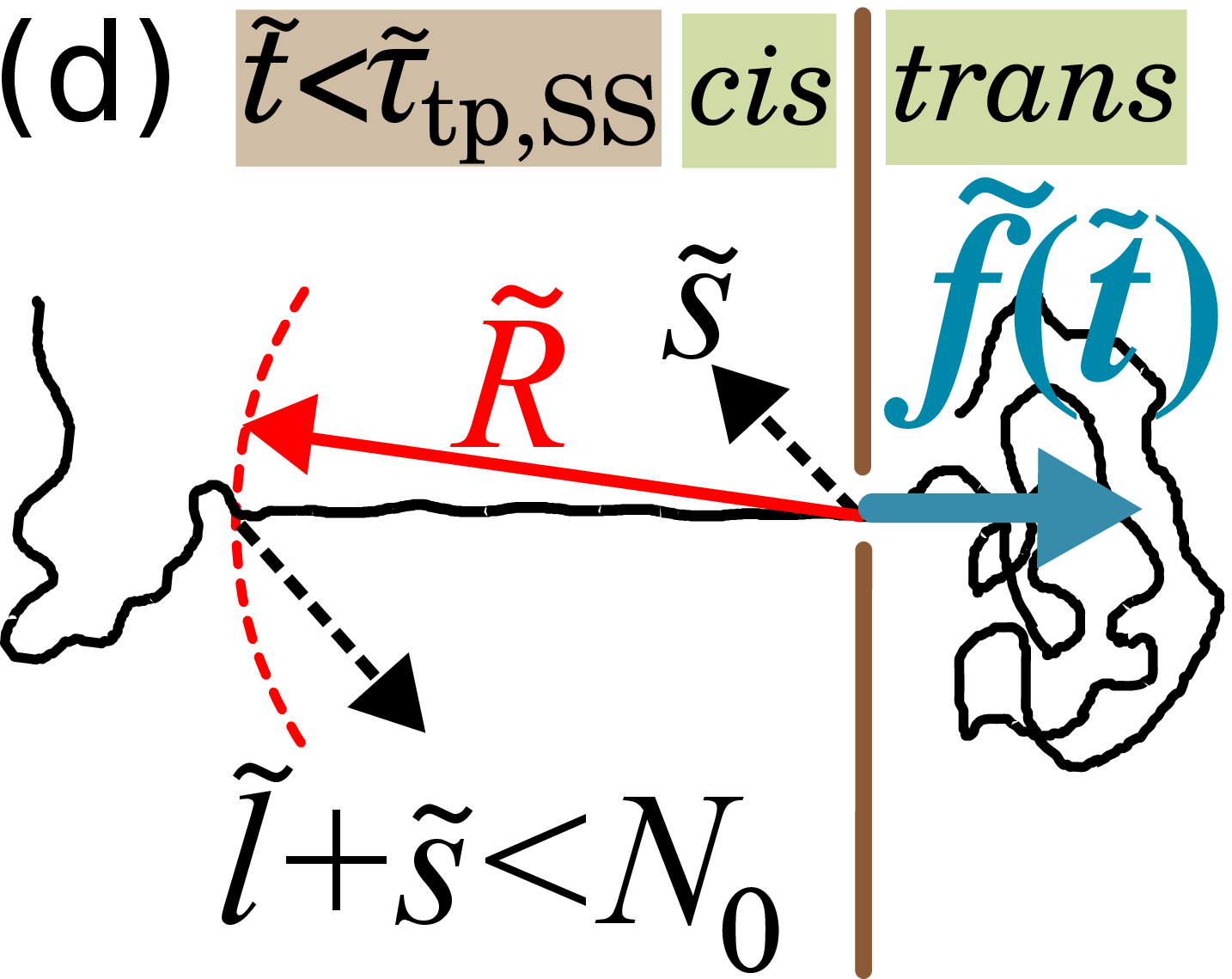}
    \end{minipage} \hskip0cm
\caption{
(a) A schematic of the translocation process in the propagation stage, i.e. $\tilde{t} < \tilde{t}_{\textrm{tp,T}}$, 
for the trumpet regime. The time dependent external driving force $ \tilde{f} ( \tilde{t} )$ acts only on the polymer bead 
in the pore towards the {\it trans} side. $N_0$ is the length of polymer and translocation coordinate, $\tilde{s}$, is 
the number of beads that have already been translocated into the {\it trans} side. $\tilde{l} + \tilde{s}$ is the number 
of beads influenced by the tension force on the {\it cis} side, which is less than $N_0$ during the propagation stage. 
The location of the tension front is determined by $\tilde{R}$. $\tilde{d} (\tilde{t})$ is the pore width that is varying in time 
during the translocation process when the pore is flickering. 
(b) The translocation process for the trumpet regime during the post propagation stage that is the regime starting after 
which the tension front reaches the chain end. Panels (c) and (d) are the same as panel (a) but for the
stem-flower and strong-stretching regimes, respectively. The propagation times for the trumpet, stem-flower and strong-stretching regimes
are defined by $\tilde{\tau}_{\textrm{tp,T}}$, $\tilde{\tau}_{\textrm{tp,SF}}$ and $\tilde{\tau}_{\textrm{tp,SS}}$, respectively.
} 
\label{fig:schimatic}
\end{center}
\end{figure*}

Polymer translocation through a nanopore, which has rapidly emerged as one of the most active research areas in biological and 
soft matter physics, plays an important role in many biological processes such as RNA transport through a nuclear 
membrane pore and $\alpha$-hemolysin channels \cite{salman_pnas_2001,kasi1996}, virus injection into cells and 
transportation of protein through membrane channels \cite{albert_book}. It also has many technological applications 
such as rapid DNA sequencing 
\cite{kasi1996,meller2003,mellerPRL2001,tapioPRL2008DNAsequencing,aksimentievNanolett2008,golestanianPRX2012}, 
drug delivery \cite{meller2003} and gene therapy, and has been motivation for many 
theoretical and experimental studies~\cite{sung1996,muthu1999,meller2003,chuang2001,metzler2003,kantor2004,storm2005,
grosberg2006,dubbeldam2007,sakaue2007,sakaue2008,luo2008,branton2008,luo2009,schadt2010,sakaue2010,
Milchev_JPCM, rowghanian2011,Muthukumar_book,ikonen2012b,ikonen2013,Tapio_review,bhatta2009,lehtola2009,bhatta2010,
lehtola2010,dubbeldam2012,saito2012,saito2012a,saito2012b,ikonen2012a}.
So far most of the theoretical studies have been focused on the case of pore-driven translocation of a 
polymer chain through a nanopore with a constant radius and by a constant driving force, which is inherently a far-from-equilibrium 
phenomenon~\cite{dubbeldam2012,saito2012,saito2012a,saito2012b,ikonen2012a,ikonen2013,bhatta2009,bhatta2010,lehtola2009,lehtola2010}.
However, some theoretical and experimental work has also been done where the width of the pore changes during the translocation process.
For example, the active translocation of a polymer chain through a flickering pore has been investigated by means of 
molecular dynamics simulations to show that translocation through the pore with an alternating width and sticky 
walls exhibits more efficient translocation as compared to the static pore \cite{golestanianPRL2011}. 
There are some biological examples of this such as the nuclear pore complex, 
which plays an important role in nucleocytoplasmic transport in eukaryotes 
\cite{yamadaFlickeringPore}. Another example is the exchange of molecules between mitochondria and the rest of cell
which is controlled by the twin-pore protein translocase (TIM22 complex) in the inner membrane of 
the mitocondria \cite{RehlingFlickeringPore}. On the experimental side, it has been shown that the translocation of DNA 
through a nanochannel can be modulated by dynamically changing the cross section of an elastomeric nanochannel device by 
applying mechanical stress \cite{HuhNature2007,AngeliLabonChip2011,fanzioSciRep2012}.
Moreover, the width of the nanopore can be tuned by using thermally driven nanoactuation of the polyNIPAM 
brushes inside the nanopore \cite{YameenSmall2009}.

The external driving force can also be time-dependent during the translocation process
\cite{aksimentievNanolett2008,ikonen2012c,golestanianJCP2012,fiasconaroPRE2015,StefureacOscillatingForce,BatesBioPhysJ2003}.
As an example, polymer translocation has been studied for a time-dependent alternating driving force by means of 
Langevin dynamics simulations. It has been shown that when the polymer-pore interaction is attractive, resonant activation 
occurs at an optimal frequency of the alternating driving force \cite{ikonen2012c}.
There are some biological applications for the alternating driving force case such as translocation of linear and $\alpha$-helical 
peptides through an $\alpha$-hemolysin pore in the presence of an AC field \cite{StefureacOscillatingForce}, and monitoring 
the escape of the DNA from an $\alpha$-hemolysin pore by using an alternating current signal \cite{lathropJACS2010OscillatingForce}.
Moreover, using an alternating electric field in the nanopore has been suggested as a method for DNA sequencing 
\cite{aksimentievNanolett2008}.

On the theoretical side, over the last few years a consistent theory of driven translocation has been developed based
on the idea of {\it tension propagation} (TP) \cite{sakaue2007}. Within the TP theory, the dynamics of driven translocation
can be described by considering a tension force that propagates along the backbone of the chain. The
dynamics is dominated by the drag of the {\it cis} side of the polymer chain, leading to asymptotic scaling
of the average translocation time
$\tau \propto N_0^{1+\nu}$ as a function of the chain length (number of monomers) $N_0$ 
\cite{ikonen2012a,ikonen2013,ikonen2012b,jalal2014}, where the Flory exponent $\nu$ is exactly $3/4$ in 
two and approximately $0.59$ in three dimensions, respectively \cite{Edwardsbook}. In addition, the theory reveals that
the pore friction  $\eta_{\textrm{p}}$ plays an important role for chains of finite length, leading to a large
correction-to-scaling term such that the translocation time actually scales as 
$\tau(N_0) = c_1 N_0^{1+\nu} +c_2 \eta_{\textrm{p}} N_0$, 
where $c_1$ and $c_2$ are constants \cite{ikonen2013,jalal2014}.
This scaling form has most recently been theoretically derived using
the iso-flux Brownian dynamics tension propagation (IFTP) theory of driven polymer 
translocation \cite{jalal2014,rowghanian2011}.

In this paper, our main goal is to investigate the influence of an alternating external driving force and a
flickering pore on the dynamics of the driven translocation process. To this end we employ the IFTP theory
to theoretically study polymer translocation in the presence of a time dependent driving force through a flickering nanopore. 
Here, the time dependent driving force, $f (t)$, is incorporated into the theory directly, while the effect of the
flickering pore is implemented by a time dependent pore friction coefficient $\eta_{\textrm{p}} (t)$. 
Using the IFTP theory modified in this manner allows us to explicitly derive the TP equations for the present setup.
The formalism also allows us to investigate both the influence of thermal fluctuations from the solvent by introducing
randomness into the effective driving force~\cite{ikonen2012a,ikonen2012b,jalal2014} and the influence of
the distribution of the initial configurations of the chain on the {\it cis} side~\cite{jalal2014}.

The outline of the paper is as follows: In Sec.~\!\ref{model} we show how to model pore-driven polymer translocation 
in the context of the iso-flux Brownian dynamics tension propagation formalism when the pore friction is time dependent 
and the driving force is alternating. Secs.~\!\ref{scaling_oscillating_force} and \ref{scaling_flickering_pore} present 
the finite-size scaling form for the translocation time in the presence of an alternating driving force and time dependent 
pore friction, respectively. Secs.~\!\ref{average_translocation_time} and \ref{Sec_Dynamics_trans_proc}
are devoted to the results on the average translocation time, the waiting time distribution, and time evolution of 
the translocation coordinate, respectively. Conclusions and discussion are in Sec.~\!\ref{conclusions}.

\section{Model} \label{model}

For the sake of brevity, we use dimensionless units denoted by tilde 
as $\tilde{Z} \equiv Z / Z_u$, with the units of length $s_u \equiv a$,
time $t_u \equiv \eta a^2 / (k_B T)$, velocity $v_u \equiv a/t_u = k_B T/(\eta a)$, force $f_u \equiv k_B T/a$, 
monomer flux $\phi_u \equiv k_B T/(\eta a^2)$, and friction $\Gamma_u \equiv \eta$,
where $a$ is the segment length, $\eta$ is the solvent friction per monomer, $k_B$ is the Boltzmann constant, 
and $T$ is the temperature of the system.

To elucidate the parameter values chosen in this paper, we compare them with those in real physical systems 
and also with those used in molecular dynamics (MD) simulations in reduced Lennard-Jones (LJ) units \cite{ikonen2012c,jalal2014}.
Here, parameters without tilde such as external driving force, $F$, pore friction, $\eta_{\textrm{p}}$, and time, $t$, are in
LJ units. In MD simulations the scale of time, mass, and energy can be fixed by the LJ parameters 
$m$, $\sigma$ and $\epsilon$ which are the mass and diameter of each bead, and the strength of the interaction, respectively.
We have chosen $k_B T =1.2$ which is dimensionless and the time is scaled by $t_{\textrm{LJ}}= (m\sigma^2 / \epsilon)^{1/2}$. 
In our model the interaction strength is $3.39 \times 10^{-21}~ \textrm{J}$ at room temperature ($T= 295 ~ \textrm{K}$), the 
mass of each bead is about 936 amu, and the size of each bead corresponds to the single-stranded DNA Kuhn length, which is 
$\sigma \approx 1.5~ {\textrm{nm}}$. Therefore, the LJ time scale is 32.1 ps. Here the thickness of the pore is set to $\sigma$. 
By assuming the effective charge of $0.094 e$ for each unit charge, three unit charges per bead \cite{brantonPRL2003,mellerbiophysj2004}, 
and with a force scale of $2.3~\textrm{pN}$, an external driving force of $F=5$ corresponds to a voltage of $375~\textrm{mV}$ 
across the pore \cite{ikonen2012c}.

We use Brownian dynamics (BD) in the overdamped limit as the basic theoretical framework \cite{ikonen2012a,ikonen2012b}. 
The time evolution of the translocation coordinate $\tilde{s}$ which is the chain length on the {\it trans} 
side is governed by the BD equation that can be written as
\begin{equation}
\tilde{\Gamma} (\tilde{t}) \frac{d \tilde{s}}{ d \tilde{t}} =
(1- \gamma ') \bigg[ \frac{1}{N_0 - \tilde{s}} - \frac{1}{\tilde{s}} \bigg] 
+ \tilde{f} (\tilde{t}) + \tilde{\zeta} (\tilde{t}) \equiv  \tilde{f}_{\textrm{tot}},
\label{BD_equation}
\end{equation}
where $\tilde{\Gamma} (\tilde{t})$ is the effective friction, the first term in the right hand side is the entropic force 
and $\gamma '$ is the surface exponent ($\gamma '  \approx ~  0.95, \approx~ 0.69$ for self-avoiding chains in two and three 
dimensions, respectively, and $\gamma '= 0.5$ for ideal chains), $N_0$ is the chain length or the total number of 
beads in the chain (here the contour length of the polymer chain is $L= a N_0$), $\tilde{f} (\tilde{t})$ 
is the external driving force, $\tilde{f}_{\textrm{tot}}$ is the total force, and the Gaussian white noise, 
$\tilde{\zeta} (\tilde{t})$, satisfies $\langle \zeta (t) \rangle = 0$ and 
$\langle \zeta (t) \zeta (t') \rangle = 2 \Gamma (t) k_B T \delta (t - t ')$.

As both the {\it cis} side subchain and the pore friction contribute to the effective friction 
$\tilde{\Gamma} (\tilde{t})$, it can be written as 
$\tilde{\Gamma} (\tilde{t}) = \tilde{\eta}_{cis} (\tilde{t}) + \tilde{\eta}_{\textrm{p}} (\tilde{t})$.
For a fully flexible self-avoiding polymer chain, the dynamical contribution of the {\it trans} side to the
dynamics has been shown to be insignificant~\cite{ikonen2012a,ikonen2012b,ikonen2013,dubbeldam2014,suhonen2014}
and it can be absorbed into the pore friction $\tilde{\eta}_{\textrm{p}} (\tilde{t})$. 
Note that here the pore friction $\tilde{\eta}_{\textrm{p}} (\tilde{t})$ is assumed to be a function of time.
We solve the dynamics of the {\it cis} side by using the TP equations.
Following the arguments of Rowghanian {\it et al.}~\cite{rowghanian2011} we are able to derive the TP equations analytically.
We assume that the monomer flux, $\tilde{\phi}\equiv d\tilde{s}/d\tilde{t}$, is constant in space on the mobile 
domain of the {\it cis} side and through the pore, but evolves in time. The tension front which is the boundary between 
the immobile and mobile domains, is located at the distance $\tilde{x}=-\tilde{R}(\tilde{t})$ from the pore. 
The external driving force inside the mobile domain is mediated by the backbone of the chain from the pore located 
at $\tilde{x}=0$ to the last mobile monomer $N$ which is located at the tension front. Using the force-balance 
relation for the differential element $d \tilde{x}$ located between $\tilde{x}$ and $\tilde{x} + d\tilde{x}$,
the magnitude of the tension force at the distance $\tilde{x}$ from the pore can be obtained. 
One can integrate this force-balance relation over the distance from the entrance of the pore to $\tilde{x}$,
to obtain the tension force at $\tilde{x}$ as $\tilde{f}(\tilde{x},\tilde{t}) = \tilde{f}_0 - \tilde{\phi} (\tilde{t}) \tilde{x}$,
where $ \tilde{f}_0 \equiv \tilde{f}_{\textrm{tot}} - \tilde{\eta}_{\textrm{p}} (\tilde{t}) ~ \tilde{\phi}(\tilde{t})$ is 
the force at the entrance of the pore (see Appendix A of Ref.~\!\citenum{jalal2014} for technical details).
As the drag of all the preceding monomers diminishes, the mediated force closer to the the tension front becomes smaller
and it vanishes at the tension front.

According to the blob theory, the shape of the mobile part of the chain can be classified into three different regimes.
For a moderate external driving force, i.e. $N_0^{-\nu} \ll \tilde{f}_0 \ll 1$, the monomer number density close to 
the pore is greater than unity, and the chain shape resembles a trumpet-like shape with the narrow end closer to the 
pore. This regime is called {\it trumpet} (TR) which is depicted in Figs.~\ref{fig:schimatic}a and b. For a
stronger external driving force, $1 \ll \tilde{f}_0 \ll N_0^{\nu}$, the force can only straighten a small part of the 
chain which is called the stem, and the part following it is called the flower. This regime is called {\it stem-flower} 
(SF) (see Fig.~\ref{fig:schimatic}c). Finally, for a very strong driving force, i.e. $\tilde{f}_0 \gg N_0^{\nu}$, 
the force is strong enough to straighten the whole moving part of the chain, which is called the {\it strong-stretching} (SS) regime 
(see Fig.~\ref{fig:schimatic}d). In the TR and SF regimes, the location 
of the tension front is at the farthest blob from the pore, while in the latter one, i.e. SS, the tension 
front is located at the farthest mobile bead of the chain from the pore as can be seen in Fig.~\ref{fig:schimatic}.

The monomer flux can be written as a function of the total force and linear size of the mobile part by
integrating the force balance equation over the whole mobile part:
\begin{equation}
\tilde{\phi} (\tilde{t}) = \frac{\tilde{f}_{\textrm{tot}} (\tilde{t})}
{\tilde{\eta}_{\textrm{p}} (\tilde{t}) + \tilde{R} (\tilde{t}) }.
\label{phi_equation}
\end{equation}
The effective friction can be obtained as
\begin{equation}
\tilde{\Gamma} (\tilde{t}) = \tilde{R}(\tilde{t}) + \tilde{\eta}_{\textrm{p}} (\tilde{t}),
\label{Gamma_equation}
\end{equation}
by combining the definition of the flux, $\tilde{\phi}\equiv d\tilde{s}/d\tilde{t}$,
with Eqs.~\!\eqref{BD_equation} and \eqref{phi_equation}.

Equations ~\!\eqref{BD_equation}, \eqref{phi_equation} and \eqref{Gamma_equation}  together with the knowledge of 
$\tilde{R}(\tilde{t})$ determine the full solution for $\tilde{s}$. Therefore, the only equation remaining 
is the equation of motion for $\tilde{R}(\tilde{t})$ that must be derived separately for the {\it propagation} 
and {\it post propagation} stages. In the propagation stage, as shown in Fig.~\ref{fig:schimatic}a, c and d, 
the tension has not reached the last monomer of the chain, i.e. $N = \tilde{l} + \tilde{s} < N_0$, where $N$ is 
the number of total beads that have been already influenced by the tension force and $\tilde{l}$ is the number 
of dragged monomers in the {\it cis} side. Using the scaling relation of the end-to-end distance of the self-avoiding 
chain, $\tilde{R} = \tilde{A}_{\nu} N^{\nu}$, where $\tilde{A}_\nu$ is a constant prefactor and $N$ is the last monomer inside 
the mobile domain at the tension front, together with the blob theory and $N = \tilde{l} + \tilde{s}$ (see 
Appendix B of Ref.~\!\citenum{jalal2014} for details) the equation of motion for the tension front in the propagation 
stage can be derived as
\begin{equation}
\dot{\tilde{R}} (\tilde{t}) \!=\!
\frac{\nu \tilde{A}_{\nu}^{ \frac{1}{\nu} }  \tilde{R} (\tilde{t})^{ \frac{\nu -1}{\nu} }
 \tilde{\phi} (\tilde{t}) }
{ 1 - \nu \tilde{A}_{\nu}^{ \frac{1}{\nu} }  \tilde{R} (\tilde{t})^{ \frac{\nu -1}{\nu} } }  ,
\label{evolution_of_R_propagation_SS}
\end{equation}
for the SS regime, and
\begin{equation}
\dot{\tilde{R}} (\tilde{t}) \!=\!
\frac{\nu \tilde{A}_{\nu}^{ \frac{1}{\nu} }  \tilde{R} (\tilde{t})^{ \frac{\nu -1}{\nu} }
\big\{  (\tilde{{\cal{L}}}_{\textrm{a}} \!+\! \tilde{{\cal{G}}}_{\textrm{a}})
\big[ \dot{\tilde{f} }_{\textrm{tot}}  (\tilde{t}) \!-\! 
\tilde{\phi} (\tilde{t}) \dot{\tilde{\eta}}_{\textrm{p}} (\tilde{t}) \big] \!\!+\! \tilde{\phi} (\tilde{t}) \big\} }
{ 1 + \nu \tilde{A}_{\nu}^{ \frac{1}{\nu} }  \tilde{R} (\tilde{t})^{ \frac{\nu -1}{\nu} }  
\tilde{{\cal{L}}}_{\textrm{a}} \times \tilde{\phi} (\tilde{t}) },
\label{evolution_of_R_propagation}
\end{equation}
for the TR and SF regimes. Here the subscript "$\textrm{a}$" in $\tilde{{\cal{L}}}_{\textrm{a}}$ and $\tilde{{\cal{G}}}_{\textrm{a}}$ 
stands for the trumpet regime as $\textrm{TR}_+$ and $\textrm{TR}_-$ correspond to positive and negative values of $\tilde{\phi}$, 
respectively. For the stem-flower regime "$\textrm{a}$" denotes $\textrm{SF}$. The time derivative of $\tilde{f}_{\textrm{tot}}$
is $\dot{\tilde{f} }_{\textrm{tot}}$. $\tilde{{\cal{L}}}_{\textrm{a}}$ and $\tilde{{\cal{G}}}_{\textrm{a}}$ are given by
\begin{subequations}
\begin{align}
\tilde{{\cal{L}}}_{\textrm{TR}_{+}} &= \frac{ 1 }{ \tilde{\eta}_{\textrm{p}} (\tilde{t}) + \tilde{R} (\tilde{t}) }
\bigg\{ -\frac{\nu}{(2\nu-1) \tilde{\phi} (\tilde{t})^2 } 
\big[ \tilde{\phi} (\tilde{t}) \tilde{R} (\tilde{t}) \big]^{\frac{2\nu-1}{\nu}} \nonumber\\
&- \frac{  \tilde{\eta}_{\textrm{p}} (\tilde{t}) }{  \tilde{\phi} (\tilde{t})  } 
\big[ \tilde{\phi} (\tilde{t}) \tilde{R} (\tilde{t})  \big]^{\frac{\nu-1}{\nu}}
\bigg\}
\hspace{+0.5cm}:~\tilde{\phi} (\tilde{t}) >0,
\\
\tilde{{\cal{L}}}_{\textrm{TR}_{-}} &= \frac{1}{\tilde{\eta}_{\textrm{p}} (\tilde{t}) + \tilde{R} (\tilde{t}) } 
\bigg\{   \frac{\nu}{(2\nu-1) \tilde{\phi} (\tilde{t})^2 } 
\big[ -\tilde{\phi} (\tilde{t}) \tilde{R} (\tilde{t}) \big]^{\frac{2\nu-1}{\nu}} \nonumber\\
&- \frac{  \tilde{\eta}_{\textrm{p}} (\tilde{t}) }{  \tilde{\phi} (\tilde{t})  } 
\big[ - \tilde{\phi} (\tilde{t}) \tilde{R} (\tilde{t}) \big]^{\frac{\nu-1}{\nu}} \bigg\}
\hspace{+0.1cm}:~\tilde{\phi} (\tilde{t}) <0,
\\
\tilde{{\cal{G}}}_{\textrm{TR}_{+}} &= \frac{ 1 }{ \tilde{\phi} (\tilde{t}) } 
\big[ \tilde{\phi} (\tilde{t}) \tilde{R} (\tilde{t}) \big]^{\frac{\nu-1}{\nu}}
\hspace{+0.8cm}:~\tilde{\phi} (\tilde{t}) >0,
\\
\tilde{{\cal{G}}}_{\textrm{TR}_{-}} &= \frac{ 1 }{ \tilde{\phi} (\tilde{t}) } 
\big[ - \tilde{\phi} (\tilde{t}) \tilde{R} (\tilde{t}) \big]^{\frac{\nu-1}{\nu}}
\hspace{+0.4cm}:~\tilde{\phi} (\tilde{t}) <0,
\\
\tilde{{\cal{L}}}_{\textrm{SF}} &=
-\frac{1}{ \tilde{\phi} (\tilde{t}) }  + 
\frac{\nu-1}{(2\nu-1) \big[ \tilde{\eta}_{\textrm{p}} (\tilde{t}) + \tilde{R} (\tilde{t}) \big] \tilde{\phi} (\tilde{t})^2 } 
, \\
\tilde{{\cal{G}}}_{\textrm{SF}} &= \frac{1}{\tilde{\phi} (\tilde{t}) }.
\end{align} 
\label{L_G}
\end{subequations}
In the post propagation stage which is schematically shown in Fig.~\ref{fig:schimatic}b for the TR regime, 
all monomers on the {\it cis} side have already been affected by the tension force. Therefore, the correct closure 
relation for the post propagation stage is $\tilde{l}+\tilde{s}=N_0$. Therefore, one can derive the equation 
of motion for the tension front as
\begin{equation}
\dot{\tilde{R}} (\tilde{t}) = - \tilde{\phi} (\tilde{t}),
\label{evolution_of_R_post_propagation_SS}
\end{equation}
for the SS regime, while for the TR and SF regimes it can be written as
\begin{equation}
\dot{\tilde{R}} (\tilde{t}) = \frac{ (\tilde{{\cal{L}}}_{\textrm{a}} + \tilde{{\cal{G}}}_{\textrm{a}}) 
\big[ \dot{\tilde{f} }_{\textrm{tot}}  (\tilde{t}) - 
\tilde{\phi} (\tilde{t}) \dot{\tilde{\eta}}_{\textrm{p}} (\tilde{t}) \big] + \tilde{\phi} (\tilde{t})}
{ \tilde{\phi} (\tilde{t}) \times \tilde{{\cal{L}}}_{\textrm{a}} }.
\label{evolution_of_R_post_propagation_2}
\end{equation}
For more details, the interested reader is referred to the Appendix B of Ref.~\!\citenum{jalal2014}.

Eqs.~\!(\ref{BD_equation}), (\ref{phi_equation}), (\ref{Gamma_equation}), (\ref{evolution_of_R_propagation_SS}) and 
(\ref{evolution_of_R_propagation}) must be solved self-consistently to find the solution for the model 
in the propagation stage, while in the post propagation stage, one must solve the set of 
Eqs.~\!(\ref{BD_equation}), (\ref{phi_equation}), (\ref{Gamma_equation}), (\ref{evolution_of_R_post_propagation_SS}) and 
(\ref{evolution_of_R_post_propagation_2}).

\section{Scaling of translocation time for SS regime} \label{scaling}
\subsection{Static pore and oscillating force} \label{scaling_oscillating_force}

\begin{figure}[t]
\begin{center}
  \includegraphics[width=0.48\textwidth]{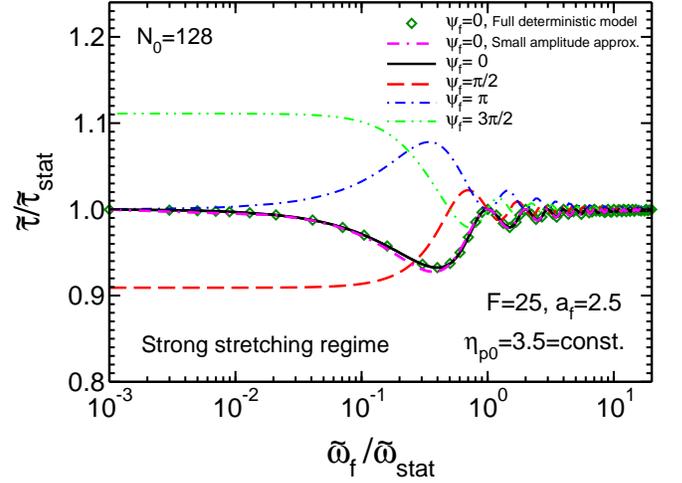}
\caption{Analytic approximation for the normalized translocation time, 
$\tilde{\tau} / \tilde{\tau}_{\textrm{stat}}$, as a function of normalized force frequency, 
$\tilde{\omega}_f / \tilde{\omega}_{\textrm{stat}}$ for various values of initial phases of 
$\psi_f= 0, \pi/2, \pi$ and $3\pi/2$ in the SS regime. Here, the pore is static with constant 
pore friction $\eta_{\textrm{p}0}= 3.5$, the chain length is $N_0=128$, and the time dependent 
external driving force is 
$\tilde{f} (\tilde{t}) = \tilde{F} + \tilde{a}_{f} \sin (\tilde{\omega}_f \tilde{t} +\psi_f )$, 
where $F= 25$ and $a_f=2.5$. The green diamonds show the normalized translocation time for the 
full IFTP model which includes the entropic force term. The magenta dash-dash-dotted curve 
presents the normalized translocation time in the small amplitude approximation.}
\label{fig:translocation_time_alternating_force_SS}
\end{center}
\end{figure}

To obtain analytical results, we assume that only the external driving force has contribution to the total force 
in the BD equation~\eqref{BD_equation}, i.e. we ignore both the entropic and thermal noise terms here. We 
choose the external driving force as a combination of a constant force $\tilde{F}$ and an oscillatory term 
$\tilde{A}_f (\tilde{t})$ as
\begin{equation}
\tilde{f}_{\textrm{tot}} = \tilde{f} (\tilde{t}) = \tilde{F} + \tilde{A}_f (\tilde{t}),
\label{oscillatory_force}
\end{equation}
where 
$\tilde{A}_f (\tilde{t}) = 
\tilde{a}_{f} \sin (\tilde{\omega}_{f} \tilde{t} +\psi_{f} )$,
and $\tilde{a}_{f}$, $\tilde{\omega}_{f}$ and $\psi_{f}$ are the amplitude, frequency and initial phase of 
the force, respectively. For simplicity, we first consider a static pore with a constant pore friction, i.e. 
$\tilde{\eta}_{\textrm{p}} (\tilde{t}) = \tilde{\eta}_{\textrm{p}0}$.
Then Eq.~\!\eqref{phi_equation} reduces to $\tilde{\phi}(\tilde{t})= \big[ \tilde{F} + \tilde{A}_f (\tilde{t}) \big] / 
\big[ \tilde{R}(\tilde{t}) + \tilde{\eta}_{\textrm{p}0} \big]$. In the strong-stretching regime considered 
here, the number of mobile monomers on the {\it cis} side is given by $\tilde{l}_{\textrm{SS}} = \tilde{R}$. 
Knowing the external driving force as a function of time and pore friction together with the conservation of 
mass, $N= \tilde{s}+ \tilde{l}$, the propagation time $\tilde{\tau}_\mathrm{tp}$ can be solved by integration 
of $N$ from $0$ to $N_0$. The result is
\begin{equation}
\tilde{\tau}_\mathrm{tp} = \frac{1}{\tilde{F}} \bigg[ \int_0^{N_0} \tilde{R}(N)dN + \tilde{\eta}_{\textrm{p}0} N_0 \bigg]
- \Delta \tilde{\tau}_\mathrm{tp},
\label{propagation_time_1}
\end{equation}
where
\begin{equation}
\Delta \tilde{\tau}_\mathrm{tp} \!\!=\!\! 
\frac{1}{\tilde{F}} \left[ \frac{1}{2} \tilde{R}^2(N_0) 
\!+\! \tilde{\eta}_{\textrm{p}0} \tilde{R}(N_0) \!+\! \int_{0}^{\tilde{\tau}_{\textrm{tp}}} 
\!\!\! \tilde{A}_f (\tilde{t}) d\tilde{t} \right]\!.
\label{propagation_time_2}
\end{equation}
In the post propagation stage, one sets the condition $dN/d\tilde{t} = 0$ and integrates $\tilde{R}$ 
from $\tilde{R}(N_0)$ to $0$ to obtain the post-propagation time, $\tilde{\tau}_\mathrm{pp}$, as
\begin{equation}
\tilde{\tau}_\mathrm{pp} = \frac{1}{\tilde{F}} \left[ \frac{1}{2} \tilde{R}^2(N_0) 
\!+\! \tilde{\eta}_{\textrm{p}0} \tilde{R}(N_0) 
\!-\! \int_{\tilde{\tau}_{\textrm{tp}}}^{\tilde{\tau}}\!\!\! \tilde{A}_f (\tilde{t}) d\tilde{t} \right]\!.
\label{scaling_trans_time_SS_1}
\end{equation}
Finally, the total translocation time is then given by
\begin{align}
\tilde{\tau} &= \tilde{\tau}_\mathrm{tp} + \tilde{\tau}_\mathrm{pp} \notag \\ 
& = \frac{1}{\tilde{F}}\left[ \int_0^{N_0} \tilde{R}(N)dN + \tilde{\eta}_{\textrm{p}0} N_0 
\!-\! \int_{0}^{\tilde{\tau}}\!\!\! \tilde{A}_f (\tilde{t}) d\tilde{t} \right] \notag \\
& = \tilde{\tau}_{\textrm{stat}}
\!-\! \frac{1}{\tilde{F}} \int_{0}^{\tilde{\tau}}\!\!\! \tilde{A}_f (\tilde{t}) d\tilde{t}.
\label{scaling_trans_time_SS_2}
\end{align}
where
\begin{equation}
\tilde{\tau}_{\textrm{stat}} = \frac{\tilde{A}_\nu}{(1+\nu)\tilde{F}} N_0^{1+\nu}+ 
\frac{\tilde{\eta}_{\textrm{p}0}}{\tilde{F}} N_0 ,
\label{scaling_trans_time_SS_3}
\end{equation}
is the translocation time in the absence of the oscillatory term in the external driving force.
 Here we have used the same value of the amplitude $\tilde{A}_{\nu} = 1.15$ as in Ref.~\!\citenum{ikonen2012a}.

In Fig.~\!\ref{fig:translocation_time_alternating_force_SS} we plot the translocation time from 
Eq.~\!\eqref{scaling_trans_time_SS_2} normalized with Eq.~\!\eqref{scaling_trans_time_SS_3}, as a function of the
normalized frequency $\tilde{\omega}_f/ \tilde{\omega}_{\textrm{stat}}$ for various values of $\psi_f$ in the SS
regime ($F=25$ and $a_f=2.5$). For comparison, we also show results from the full deterministic IFTP model
(green diamonds) where the entropic force term has been included. It can be seen that the agreement
with the analytic approximation in the SS regime is very good for a wide range of frequencies.

Using the explicit form of 
$\tilde{A}_f (\tilde{t})= \tilde{a}_{f} \sin (\tilde{\omega}_{f} \tilde{t} +\psi_{f} )$ we can now examine
Eq.~\!\eqref{scaling_trans_time_SS_2} in different limits. First, in
the high frequency limit, i.e. $\tilde{\omega}_{f} \gg \tilde{\tau}^{-1}$, the integral
$\int_{0}^{\tilde{\tau}}\!\!\! \tilde{A}_f (\tilde{t}) d\tilde{t}$ simply
vanishes and the total translocation time in the presence of the oscillatory term is the same as the 
translocation time in the absence of it, i.e. $\tilde{\tau} = \tilde{\tau}_{\textrm{stat}}$, as
can also be seen in the high-frequency limit of Fig.~\!\ref{fig:translocation_time_alternating_force_SS}.

In the opposite limit of a very small frequency, $\tilde{\omega}_f \ll \tilde{\tau}^{-1}$, the integral gives
\begin{eqnarray}
~\!\!\!\!\!\!\!\!\!\!\!\!-\frac{1}{\tilde{F}} \!\! \int_{0}^{\tilde{\tau}}\!\!\! \tilde{A}_f (\tilde{t}) d\tilde{t} 
&=&
-\frac{1}{\tilde{F}} \int_{0}^{\tilde{\tau}}\!\!\!
\tilde{a}_{f} \sin (\tilde{\omega}_{f} \tilde{t} +\psi_{f} ) d\tilde{t} \nonumber\\
~\!\!\!\!&=&  - \frac{\tilde{a}_f}{\tilde{F} \tilde{\omega}_f} \! \bigg[\!\! \cos (\psi_f) \!-\! 
\cos (\tilde{\omega}_f \tilde{\tau} \!+\! \psi_f) \! \bigg] .
\label{integral_A_f}
\end{eqnarray}
The expansion of $\cos (\tilde{\omega}_f \tilde{\tau} \!+\! \psi_f)$ 
 to first order yields $\cos (\tilde{\omega}_f \tilde{\tau} \!+\! \psi_f) \sim \cos (\psi_f)
- \tilde{\omega}_f \tilde{\tau} \sin (\psi_f)$ and thus
\begin{equation}
\tilde{\tau} = 
\frac{\tilde{\tau}_{\textrm{stat}}}{\big[ 1+ \tilde{a}_{f} \sin (\psi_f) / \tilde{F} \big]}.
\label{scaling_trans_time_SS_small_frequency}
\end{equation}

\begin{figure}[t]
\begin{center}
  \includegraphics[width=0.48\textwidth]{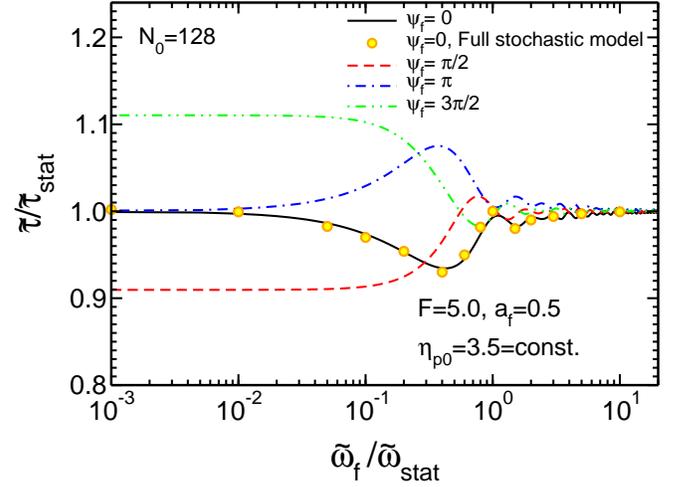}
\caption{ The normalized translocation time, $\tilde{\tau} / \tilde{\tau}_{\textrm{stat}}$, as a function of normalized force frequency, 
$\tilde{\omega}_f / \tilde{\omega}_{\textrm{stat}}$ for various values of initial phases of $\psi_f= 0, \pi/2, \pi$ and $3\pi/2$. 
Here, the pore is static with constant pore friction $\eta_{\textrm{p}0}= 3.5$, the chain 
length is $N_0=128$, and the time dependent external driving force is
$\tilde{f} (\tilde{t}) = \tilde{F} + \tilde{a}_{f} \sin (\tilde{\omega}_f \tilde{t} +\psi_f )$, where $F= 5.0$ and $a_f=0.5$. 
The orange circles show the normalized translocation time for the full model where includes thermal noise and initial 
shape distribution of the chain.}
\label{fig:translocation_time_alternating_force}
\end{center}
\end{figure}

\begin{figure}[t]
\begin{center}
  \includegraphics[width=0.48\textwidth]{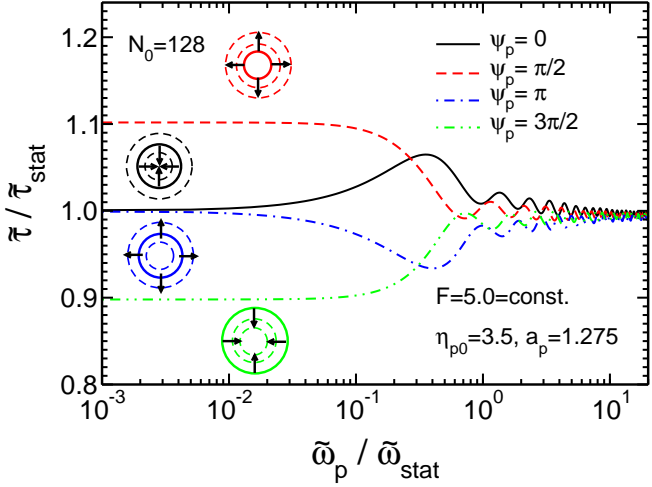}
\caption{The normalized translocation time, $\tilde{\tau} / \tilde{\tau}_{\textrm{stat}}$, as a function of normalized frequency of the 
flickering pore, $\tilde{\omega}_{\textrm{p}} / \tilde{\omega}_{\textrm{stat}}$, for various values of initial phases of 
$\psi_{\textrm{p}}= 0, \pi/2, \pi$ and $3\pi/2$. Here, the driving force is constant, $F=5.0$, the chain length 
is $N_0=128$, and the time dependent pore friction is 
$\tilde{\eta}_{\textrm{p}} (\tilde{t}) = \tilde{\eta}_{\textrm{p}0} + \tilde{a}_{\textrm{p}} 
\sin (\tilde{\omega}_{\textrm{p}} \tilde{t} +\psi_{\textrm{p}} )$, where $\eta_{\textrm{p}0}= 3.5$ and $a_{\textrm{p}}=1.275$. 
The arrows indicate which way the pore moves at the beginning of the drive, due to the initial phase $\psi_{\textrm{p}}$.}
\label{fig:translocation_time_flickering_pore}
\end{center}
\end{figure}

As can be seen from Eq.~\!\eqref{scaling_trans_time_SS_2}, the scaling form for the average translocation time is 
the same for both static and high frequency limits. In the low frequency limit, the scaling with respect to the chain 
length is also the same, but the scaling form with respect to the external driving force becomes
$\tilde{\tau} \sim [\tilde{F} + \tilde{a}_{f} \sin (\psi_f)]^{-1} $.

Finally, if the amplitude of the oscillating term, $\tilde{a}_f$, is much smaller than the static force, $\tilde{F}$, then
we can use a small amplitude approximation, where the value of the upper limit in the integral over time in 
Eq.~\!(\ref{scaling_trans_time_SS_2}) can be replaced by $\tilde{\tau}_{\textrm{stat}}$. The translocation 
time can then be written as
\begin{equation}
\tilde{\tau} = \tilde{\tau}_{\textrm{stat}} \bigg[ \!
1 + \frac{\tilde{a}_f}{2 \pi \tilde{F}} 
\frac{\cos(2\pi \frac{\tilde{\omega}_f}{\tilde{\omega}_{\textrm{stat}}} + \psi_f) 
- \cos(\psi_f)}{\frac{\tilde{\omega}_f}{\tilde{\omega}_{\textrm{stat}}}}
\bigg].
\label{scaling_trans_time_SS_small_amplitude_approximation}
\end{equation}
This equation predicts an oscillatory dependence of the translocation time on the driving frequency.
In Fig.~\!\ref{fig:translocation_time_alternating_force_SS} we plot the small amplitude expansion result 
for the case where the amplitude ratio $a_f/F=0.1$ (magenta dash-dash-dotted line).
For this relatively small ratio, the agreement with the full solution of the model (orange curve) is good for a wide
range of frequencies.

\subsection{Flickering pore and oscillating force} \label{scaling_flickering_pore}

We consider next the most general case of an oscillating force and a flickering pore in the SS regime. 
We again neglect both the entropic force and thermal noise,
and take the oscillatory force to be the same as in the previous section.
The flickering pore has the following time dependent pore friction:
\begin{equation}
\tilde{\eta}_{\textrm{p}} (\tilde{t}) = \tilde{\eta}_{\textrm{p}0} + \tilde{A}_{\textrm{p}} (\tilde{t}),
\label{flickering_pore}
\end{equation}
where $\tilde{A}_{\textrm{p}} (\tilde{t}) = \tilde{a}_{\textrm{p}} \sin (\tilde{\omega}_{\textrm{p}} \tilde{t} +\psi_{\textrm{p}} )$
and $\tilde{a}_{\textrm{p}}$, $\tilde{\omega}_{\textrm{p}}$ and $\psi_{\textrm{p}}$ are the amplitude, frequency and initial phase of the 
pore friction, respectively. 
Then, Eq.~\!\eqref{phi_equation} reduces to $\tilde{\phi}(\tilde{t})= \big[ \tilde{F} + \tilde{A}_f (\tilde{t}) \big] / 
\big[ \tilde{R}(\tilde{t}) + \tilde{\eta}_{\textrm{p}0} + \tilde{A}_{\textrm{p}} (\tilde{t}) \big]$. 
As in Sec.~\!\ref{scaling_oscillating_force}, we solve the propagation time by integration of $N$ from $0$ to $N_0$, 
and the post-propagation time by integration of $\tilde{R}$ from $\tilde{R}(N_0)$ to $0$. 
The total translocation time is then given by
\begin{align}
\tilde{\tau} 
& = \tilde{\tau}_{\textrm{stat}}
\!-\! \frac{1}{\tilde{F}} \int_{0}^{\tilde{\tau}}\!\!\! \tilde{A}_f (\tilde{t}) d\tilde{t}
+  \frac{1}{\tilde{F}} \int_{0}^{N_0} \!\!\! \tilde{A}_{\textrm{p}} (\tilde{t}) dN.
\label{scaling_trans_time_SS}
\end{align}
Using the explicit form of 
$\tilde{A}_{\textrm{p}} (\tilde{t})= \tilde{a}_{\textrm{p}} \sin (\tilde{\omega}_{\textrm{p}} \tilde{t} + \psi_{\textrm{p}} )$,
at the high frequency limit, $\tilde{\omega}_{\textrm{p}} \gg \tilde{\tau}^{-1}$, 
and for very long chains the integral
$\int_{0}^{N_0} \!\!\! \tilde{A}_{\textrm{p}} (\tilde{t}) dN$ 
vanishes and the whole translocation time in the presence of the oscillatory force
is the same as the translocation time with a constant pore friction $\tilde{\eta}_{\textrm{p}0}$. 
We note, however, that for short chains the
integral $\int_{0}^{N_0} \!\!\! \tilde{A}_{\textrm{p}} (\tilde{t}) dN$
can be non-zero due to the interaction of the polymer beads with the pore \cite{golestanianPRL2011}.

In the opposite limit of a low frequency the integral gives
\begin{eqnarray}
\!\!\!\!\!\!\!\!\!\!\frac{1}{\tilde{F}} \int_{0}^{N_0} \!\!\! \tilde{A}_{\textrm{p}} (\tilde{t}) dN \!\!&=&\!\!
\frac{1}{\tilde{F}} \int_{0}^{N_0} \!\!\! \tilde{a}_{\textrm{p}} 
\sin (\tilde{\omega}_{\textrm{p}} \tilde{t} +\psi_{\textrm{p}} ) dN\!\!\! \nonumber\\
&=& \!\!\! \frac{\tilde{a}_{\textrm{p}} \sin (\psi_{\textrm{p}} )}{\tilde{F} } N_0 .
\label{integral_A_f}
\end{eqnarray}
Therefore, 
\begin{align}
\tilde{\tau}
& = \frac{1}{\tilde{F}} \bigg[ \int_0^{N_0} \tilde{R}(N)dN 
+ \{ \tilde{\eta}_{\textrm{p}0} + \tilde{a}_{\textrm{p}} \sin (\psi_{\textrm{p}} ) \} N_0 \notag \\
& - \int_{0}^{\tilde{\tau}}\!\!\! \tilde{A}_f (\tilde{t}) d\tilde{t} 
 \bigg] = \tilde{\tau}_{\textrm{stat}} + 
\frac{1}{\tilde{F}} \bigg[ \tilde{a}_{\textrm{p}} \sin (\psi_{\textrm{p}} )N_0 
-\!\!\! \int_{0}^{\tilde{\tau}}\!\!\! \tilde{A}_f (\tilde{t}) d\tilde{t}  \bigg].
 \notag \\
\label{scaling_trans_time_SS_low_frequency_flickering}
\end{align}
As we can see in Eq.~\!(\ref{scaling_trans_time_SS_low_frequency_flickering}), at the low frequency limit the pore behaves like 
a static pore with pore friction of $\tilde{\eta}_{\textrm{p}0} + \tilde{a}_{\textrm{p}} \sin (\psi_{\textrm{p}} )$.

Equations \!\eqref{scaling_trans_time_SS} and \eqref{scaling_trans_time_SS_low_frequency_flickering} show that the scaling form
for the average translocation time for a very long chain at the high and also at the low pore frequency limits is the same 
as in the case where only the driving force is alternating and the pore is static (as discussed in the previous subsection).
For a short chain at both the high and the low pore frequency limits the scaling may be affected by the value of the integral 
$(1/{\tilde{F}}) \int_{0}^{N_0} \!\!\! \tilde{A}_{\textrm{p}} (\tilde{t}) dN$. 

When the driving force is constant, if the amplitude of the oscillating term, $\tilde{a}_p$, is much smaller 
than the static pore friction, $\tilde{\eta}_{\textrm{p}0}$, then we can also use here a small amplitude 
approximation, where the value of the time in the second integral in Eq.~\!\eqref{scaling_trans_time_SS} 
must be replaced by $\tilde{t} ( \tilde{N}(\tilde{\eta}_{\textrm{p}0}) )$. Then,
for $\tilde{A}_{\textrm{p}} (\tilde{t})= \tilde{a}_{\textrm{p}} 
\sin (\tilde{\omega}_{\textrm{p}} \tilde{t} + \psi_{\textrm{p}} )$, Eq.~\!\eqref{scaling_trans_time_SS} 
can be cast into
\begin{align}
\tilde{\tau} 
& = \tilde{\tau}_{\textrm{stat}}
+  \frac{\tilde{a}_{\textrm{p}} }{\nu \tilde{F} A_{\nu}^{1/\nu} } \int_{0}^{\tilde{R}(N_0)} 
d\tilde{R} ~ \tilde{R}^{\frac{1-\nu}{\nu}} \sin[\tilde{\omega}_{\textrm{p}} ~\tilde{t}(\tilde{R}) + \psi_{\textrm{p}}],
\label{scaling_trans_time_small-amplitude_pore-fric-SS}
\end{align}
where 
$\tilde{t}(\tilde{R}) = \frac{1}{\tilde{F}} \bigg[ \frac{ \tilde{R}^{\frac{1+\nu}{\nu}} }{ (1+\nu) A_{\nu}^{1/\nu} } 
-\frac{\tilde{R}^2}{2} + \big( \frac{ \tilde{R}^{1/\nu} }{ A_{\nu}^{1/\nu} } - \tilde{R} \big) \tilde{\eta}_{\textrm{p}0} \bigg]$.
The result of Eq.~\!\eqref{scaling_trans_time_small-amplitude_pore-fric-SS} has been checked and it matches very well 
with the result of the full deterministic model in a wide range of frequencies.

\begin{figure*}[t]
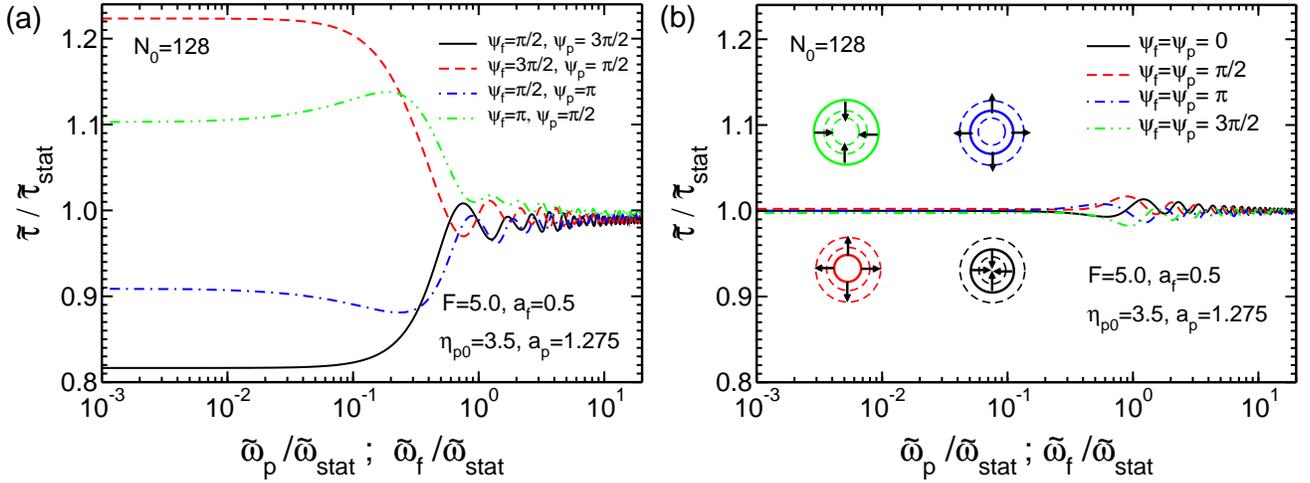
\begin{center}
    \begin{minipage}[b]{0.49\textwidth}\begin{center}
        \includegraphics[width=0.97\textwidth]{figure5a.eps}
    \end{center}\end{minipage} \hskip-0.2cm
    \begin{minipage}[b]{0.49\textwidth}\begin{center}
        \includegraphics[width=0.97\textwidth]{figure5b.eps}
    \end{center}\end{minipage} \hskip0cm
\caption{(a) The normalized translocation time, $\tilde{\tau} / \tilde{\tau}_{\textrm{stat}}$, as 
a function of either normalized pore frequency, $\tilde{\omega}_{\textrm{p}} / \tilde{\omega}_{\textrm{stat}}$, 
or force frequency, $\tilde{\omega}_f / \tilde{\omega}_{\textrm{stat}}$, for mixed initial phases 
$\psi_f= \pi/2$ and $\psi_{\textrm{p}}= 3\pi/2$, $\psi_f= 3\pi/2$ and $\psi_{\textrm{p}}= \pi/2$, 
$\psi_f= \pi/2$ and $\psi_{\textrm{p}}= \pi$, and $\psi_f= \pi$ and $\psi_{\textrm{p}}= \pi/2$.
Here, the pore is flickering and the pore friction is given by 
$\tilde{\eta}_{\textrm{p}} (\tilde{t}) = \tilde{\eta}_{\textrm{p}0} + \tilde{a}_{\textrm{p}} 
\sin (\tilde{\omega}_{\textrm{p}} \tilde{t} +\psi_{\textrm{p}} )$, where $\eta_{\textrm{p}0}=3.5$ 
and $a_{\textrm{p}}=1.275$ while the oscillating external driving force is 
$\tilde{f} (\tilde{t}) = \tilde{F} + \tilde{a}_{f} \sin (\tilde{\omega}_f \tilde{t} +\psi_f )$, 
where $F=5.0$ and $a_f=0.5$. (b) The normalized translocation time as a function of either 
$\tilde{\omega}_{\textrm{p}} / \tilde{\omega}_{\textrm{stat}} $ or $\tilde{\omega}_f / \tilde{\omega}_{\textrm{stat}}$, 
but for the identical initial pore and force phases $\psi_f= \psi_{\textrm{p}}= 0, \pi/2, \pi$ 
and $3\pi/2$, with the same parameters as of (a). The arrows show the direction of the motion of 
a point on the pore wall at the starting of the translocation process. For both panels (a) and 
(b) the chain length is $N_0=128$ and 
$\tilde{\omega}_{\textrm{p}} / \tilde{\omega}_{\textrm{stat}} = \tilde{\omega}_f / \tilde{\omega}_{\textrm{stat}}$.} 
\label{fig:rescaled_translocation_time}
\end{center}
\end{figure*}

\section{Average translocation time} \label{average_translocation_time}

\begin{figure}[t]
\begin{center}
  \includegraphics[width=0.47\textwidth]{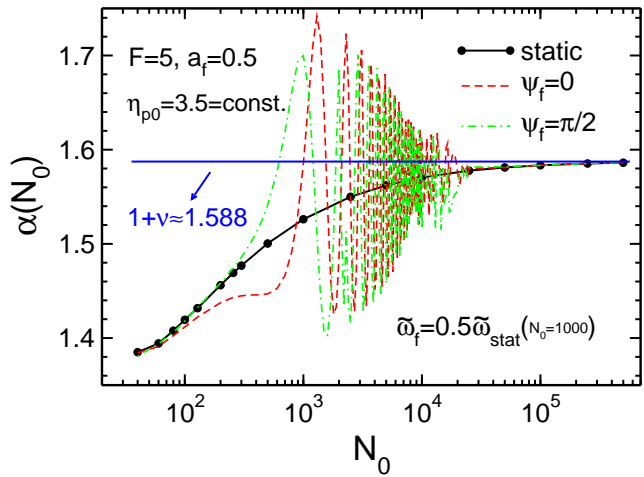}
\caption{The effective translocation exponent, $\alpha (N_0)$, as a function of the chain length, $N_0$,
for two different values of initial phases $\psi_f= 0$ (red dashed curve) and $\pi/2$ (green dashed-dotted curve).
The black solid circle-line curve shows the static case where both the pore friction and the external driving 
force are constant with the values $\eta_{\textrm{p}0}= 3.5$ and $F= 5.0$. Red and green curves are for the cases 
where the pore friction is constant, $\eta_{\textrm{p}0}= 3.5$, but the external driving force is time-dependent 
and given by $\tilde{f} (\tilde{t}) = \tilde{F} + \tilde{a}_{f} \sin (\tilde{\omega}_f \tilde{t} +\psi_f )$, 
with $F= 5.0$, $a_f=0.5$ and fixed $\tilde{\omega}_f = 0.5 \tilde{\omega}_{\textrm{stat}} (N_0=1000)= 9.5 \times 10^{-4}$. 
The horizontal blue solid line is the asymptotic value of the translocation exponent which is $1+\nu$ in the 
limit of very long chains.} 
\label{fig:translocation_exponent}
\end{center}
\end{figure}

Next we consider the average translocation time, $\tilde{\tau}$, in all the different regimes using the full IFTP theory.
To this end, we numerically solve the IFTP Eqs.~\!(\ref{BD_equation}), (\ref{phi_equation}), 
(\ref{Gamma_equation}), (\ref{evolution_of_R_propagation_SS}), (\ref{evolution_of_R_propagation}), 
(\ref{evolution_of_R_post_propagation_SS}) and (\ref{evolution_of_R_post_propagation_2}) to obtain the total translocation 
time. We study the behavior of the translocation time first for a time-dependent force, time-dependent pore friction, and 
finally when both the force and the pore friction are functions of time.

In Fig. \ref{fig:translocation_time_alternating_force} the normalized translocation time, 
$\tilde{\tau} / \tilde{\tau}_{\textrm{stat}}$, has been plotted as a function of normalized 
frequency, $\tilde{\omega}_f / \tilde{\omega}_{\textrm{stat}}$ for various values of the 
initial phases of $\psi_f= 0, \pi/2, \pi$ and $3\pi/2$. The pore is static with the 
pore friction $\eta_{\textrm{p}0}= 3.5$, chain length is $N_0=128$, and the time dependent 
external driving force is given by 
$\tilde{f} (\tilde{t}) = \tilde{F} + \tilde{a}_{f} \sin (\tilde{\omega}_f \tilde{t} +\psi_f )$, 
where $F= 5.0$ and $a_f=0.5$. Here, $\tilde{\omega}_{\textrm{stat}} = 2 \pi / \tilde{\tau}_{\textrm{stat}} $ 
is the frequency related to the whole translocation process when the pore is not flickering, 
i.e. $\tilde{\eta}_{\textrm{p}} (\tilde{t}) = \tilde{\eta}_{\textrm{p}0}$ and the external 
driving force is constant, i.e. $\tilde{f} (\tilde{t}) = \tilde{F}$. For $\psi_f= 0$ 
(black solid curve) at low frequencies the magnitude of the translocation time is the 
same as of the static force case. At this low frequency limit the translocation process 
occurs during the first half of the period of the oscillating force. During this first half 
of the cycle the value of the force increases from its static value to its maximum and back 
to the static value at $\tilde{\omega}_f / \tilde{\omega}_{\textrm{stat}}=0.5$, being always 
greater than its static value. Therefore, the translocation time gradually decreases for 
small frequencies, and has a minimum at $\tilde{\omega}_f / \tilde{\omega}_{\textrm{stat}}=0.5$.
For frequencies higher than $\tilde{\omega}_f / \tilde{\omega}_{\textrm{stat}}=0.5$, the 
chain starts experiencing the second half of the cycle where the value of the force is always 
smaller than its static limit, and thus there is a first maximum at 
$\tilde{\omega}_f / \tilde{\omega}_{\textrm{stat}}=1$. For higher frequencies, i.e. 
$\tilde{\omega}_f / \tilde{\omega}_{\textrm{stat}}> 1$, the chain again experiences the following 
half of the cycle, i.e. $\tilde{T}_f < \tilde{t} < 3\tilde{T}_f/2$ ($\tilde{T}_f = 2 \pi / \tilde{\omega}_f$), 
where again the value of the force is greater than its static limit and the translocation time 
increases. With increasing frequency the translocation time displays oscillations between minima 
and maxima with a decreasing amplitude upon approaching the high-frequency limit, where 
the rapidly oscillating force component averages to zero, as discussed in Sec.~\!\ref{scaling}. 
For comparison, we have also solved the full stochastic IFTP model for the initial phase 
$\psi_f=0$, with results shown in orange circles in Fig.~\!\ref{fig:translocation_time_alternating_force}. 
In this case we take into account both thermal noise and the initial shape distribution of the 
chain as explained in detail in Ref.~\!\citenum{jalal2014}. The results are in very good agreement 
with the deterministically solved equations.

Next, we consider the case of a flickering pore with a constant drive.
In Fig.~\!\ref{fig:translocation_time_flickering_pore} the normalized translocation time, $\tilde{\tau} / \tilde{\tau}_{\textrm{stat}}$, 
has been plotted as a function of normalized frequency of the flickering pore, $\tilde{\omega}_{\textrm{p}} / \tilde{\omega}_{\textrm{stat}}$, 
for a constant driving force, $F=5.0$, chain length of $N_0=128$, and time dependent pore friction 
$\tilde{\eta}_{\textrm{p}} (\tilde{t}) = \tilde{\eta}_{\textrm{p}0} + 
\tilde{a}_{\textrm{p}} \sin (\tilde{\omega}_{\textrm{p}} \tilde{t} +\psi_{\textrm{p}} )$ with
static pore friction $\eta_{\textrm{p}0}= 3.5$ and amplitude $a_{\textrm{p}}=1.275$, for various values of the initial phases of 
$\psi_{\textrm{p}}= 0, \pi/2, \pi$ and $3\pi/2$. 
As we can see for $\psi_{\textrm{p}}= 0$, in the low frequency limit the value of the translocation time reduces to its static 
value, where $\tilde{\tau} / \tilde{\tau}_{\textrm{stat}}  =1$. As the pore oscillation frequency increases the value of the pore friction 
also increases, 
which means that the pore width, $\tilde{d} (\tilde{t})$, decreases (see Fig.~\!\ref{fig:schimatic}a). 
Therefore, the effective interaction between the pore and the beads becomes stronger and the result is that the value of the translocation time 
grows and reaches its maximum at $\tilde{\omega}_{\textrm{p}} / \tilde{\omega}_{\textrm{stat}}=0.5$. At this frequency the translocation process 
occurs during the first half period of the pore motion. During this first half of the cycle the radius of the pore decreases from its 
static value to its minimum and back to the static limit
at $\tilde{\omega}_{\textrm{p}} / \tilde{\omega}_{\textrm{stat}}=0.5$, being always 
smaller than the radius of the static pore. Therefore, the translocation time gradually increases for low frequencies, and 
then it has a maximum at $\tilde{\omega}_{\textrm{p}} / \tilde{\omega}_{\textrm{stat}}=0.5$. For frequencies higher than 
$\tilde{\omega}_{\textrm{p}} / \tilde{\omega}_{\textrm{stat}}=0.5$, the chain experiences the second half of the cycle where the radius of the 
pore is larger than the radius of the static pore, and thus there is a strong minimum at 
$\tilde{\omega}_{\textrm{p}} / \tilde{\omega}_{\textrm{stat}}=1$. 
For even higher frequencies, i.e. $\tilde{\omega}_{\textrm{p}} / \tilde{\omega}_{\textrm{stat}}> 1$, the chain experiences the next half 
of the cycle, i.e. $\tilde{T}_{\textrm{p}} < \tilde{t} < 3\tilde{T}_{\textrm{p}}/2$ ($\tilde{T}_{\textrm{p}}= 2 \pi /\tilde{\omega}_{\textrm{p}}$), 
where the radius of the pore is again smaller than its static value and the translocation time increases. This alternation of high and low pore 
friction creates successive minima and maxima in a way similar to the case of the time-dependent drive, except that
increasing the pore friction leads to increasing translocation time, while increasing the drive leads to more rapid translocation.

Finally, we consider the combination of an oscillating drive and pore. The fact that increasing the drive or increasing friction 
work in the opposite directions with respect to the translocation time suggests that it could be minimized by out-of-phase
combination of the two forces. In Fig.~\!\ref{fig:rescaled_translocation_time}a we present the normalized translocation time as a function of 
either the normalized pore frequency or force frequency, for mixed initial phases $\psi_f= \pi/2$ and $\psi_{\textrm{p}}= 3\pi/2$,
$\psi_f= 3\pi/2$ and $\psi_{\textrm{p}}= \pi/2$, $\psi_f= \pi/2$ and $\psi_{\textrm{p}}= \pi$, and 
$\psi_f= \pi$ and $\psi_{\textrm{p}}= \pi/2$.
Here, the pore is flickering and the pore friction is given by 
$\tilde{\eta}_{\textrm{p}} (\tilde{t}) = \tilde{\eta}_{\textrm{p}0} + \tilde{a}_{\textrm{p}} 
\sin (\tilde{\omega}_{\textrm{p}} \tilde{t} +\psi_{\textrm{p}} )$, where $\eta_{\textrm{p}0}=3.5$ and $a_{\textrm{p}}=1.275$
while the oscillating external driving force is 
$\tilde{f} (\tilde{t}) = \tilde{F} + \tilde{a}_{f} \sin (\tilde{\omega}_f \tilde{t} +\psi_f )$ where $F=5.0$ and $a_f=0.5$.
The pore frequency is equal to the force frequency during the translocation process here.
For fixed values of $F$, $a_f$, $\eta_{\textrm{p}0}$ and $a_{\textrm{p}}$, the minimum value of the normalized 
translocation time, $\tilde{\tau}_{\textrm{min}} / \tilde{\tau}_{\textrm{stat}} $, occurs at 
very low frequency limit when the force phase is $\psi_f = \pi /2$ and pore phase is $\psi_{\textrm{p}} = 3 \pi /2$,
because the driving force has its maximum value and the pore friction has its minimum value.
Then the normalized translocation time gets its maximum value, $\tilde{\tau}_{\textrm{max}} / \tilde{\tau}_{\textrm{stat}} $, 
when $\psi_f = 3\pi /2$ and pore phase is $\psi_{\textrm{p}} = \pi /2$, 
because the driving force has its minimum value while the pore friction has its maximum value.
All other values of the normalized translocation times, $\tilde{\tau} / \tilde{\tau}_{\textrm{stat}}$, for the other values 
of the force and the pore phases at different force or pore frequencies are between these two extremum values, i.e. 
$\tilde{\tau}_{\textrm{min}} / \tilde{\tau}_{\textrm{stat}} 
< \tilde{\tau} / \tilde{\tau}_{\textrm{stat}} < \tilde{\tau}_{\textrm{Max}} / \tilde{\tau}_{\textrm{stat}} $.

In Fig.~\!\ref{fig:rescaled_translocation_time} b we show the same quantity as a function of either 
$\tilde{\omega}_{\textrm{p}} / \tilde{\omega}_{\textrm{stat}}$ or $\tilde{\omega}_f / \tilde{\omega}_{\textrm{stat}}$, 
but the initial pore and force 
phases are now identical, and $\psi_f= \psi_{\textrm{p}}= 0, \pi/2, \pi$ and $3\pi/2$, with the same parameters as of 
Fig.~\!\ref{fig:rescaled_translocation_time}a. Both the pore friction and the driving force are time dependent with 
the same forms and the same parameters as of Fig.~\!\ref{fig:rescaled_translocation_time}a.
As expected, the two effects now work against each other, and there's an almost complete
cancellation in the low and high frequency limits. 
As the effect of time dependent driving force and pore friction are dynamically coupled to each other, 
it is very difficult to find parameters in such a way that at low frequency limit they exactly cancel the effect of each other.
For intermediate frequencies there are some
oscillations due to the fact that the total effective friction,
Eq.~\!\eqref{Gamma_equation}, is determined dynamically by tension propagation on the {\it cis} side chain,
and cannot be completely controlled externally. In the end, adding a 
time-dependent component to the pore friction and the drive is fundamentally different.

Finally, it is interesting to see how the (effective) translocation exponent $\alpha$, defined as $\alpha (N_0) = d\ln \tilde{\tau} / d\ln N_0$, 
is affected by the presence of an alternating external driving force. For a static driving force, the correction-to-scaling term due to the
pore friction is large here and $\alpha$ slowly approaches the value 
$1+\nu \approx 1.588$ with increasing chain length\cite{ikonen2012a,ikonen2012b,ikonen2013,jalal2014}, as shown in 
Fig.~\!\ref{fig:translocation_exponent}. For the oscillating force, $\tilde{f} (\tilde{t}) = \tilde{F} + \tilde{a}_{f} 
\sin (\tilde{\omega}_f \tilde{t} +\psi_f )$ with $F= 5.0$, $a_f=0.5$, constant pore friction $\eta_{\textrm{p}0}=3.5$,
and fixed $\tilde{\omega}_f = 0.5 \tilde{\omega}_{\textrm{stat}} (N_0=1000)$, we see a very strong oscillatory behavior 
of $\alpha$. The chosen frequency $\tilde{\omega}_f = 0.5 \tilde{\omega}_{\textrm{stat}} (N_0=1000)=9.5 \times 10^{-4}$ 
is moderate compared to the translocation time of the chain of length $N_0 = 1000$, low for a short chain such as $N_0 = 40$, 
and very high for a long chain such as $N_0 = 10^ 5$. By fixing the frequency when the chain length is changed, the limits 
of low, moderate and high frequencies are all explored. At the very short and very long chain limits the translocation exponent is 
the same as for constant driving and independent of the value of the initial phase $\psi_f$. However, close to 
$N_0 = 1000$, the translocation exponent oscillates due to the oscillation of the translocation time 
(see Fig.~\!\ref{fig:translocation_time_alternating_force}). It is worth noting that while the amplitude of the exponent's oscillation 
depends on the ratio of $a_f / F$, even with a moderate value of the ratio of $a_f / F=0.1$ the exponent varies between 
$1.40$ and $1.74$, going well beyond $1+\nu$. In addition, the oscillation amplitude decreases towards long chains, as the 
system effectively enters the high frequency limit and the average driving force approaches the static force value.

\section{Dynamics of the translocation process}\label{Sec_Dynamics_trans_proc}

\begin{figure*}[t]
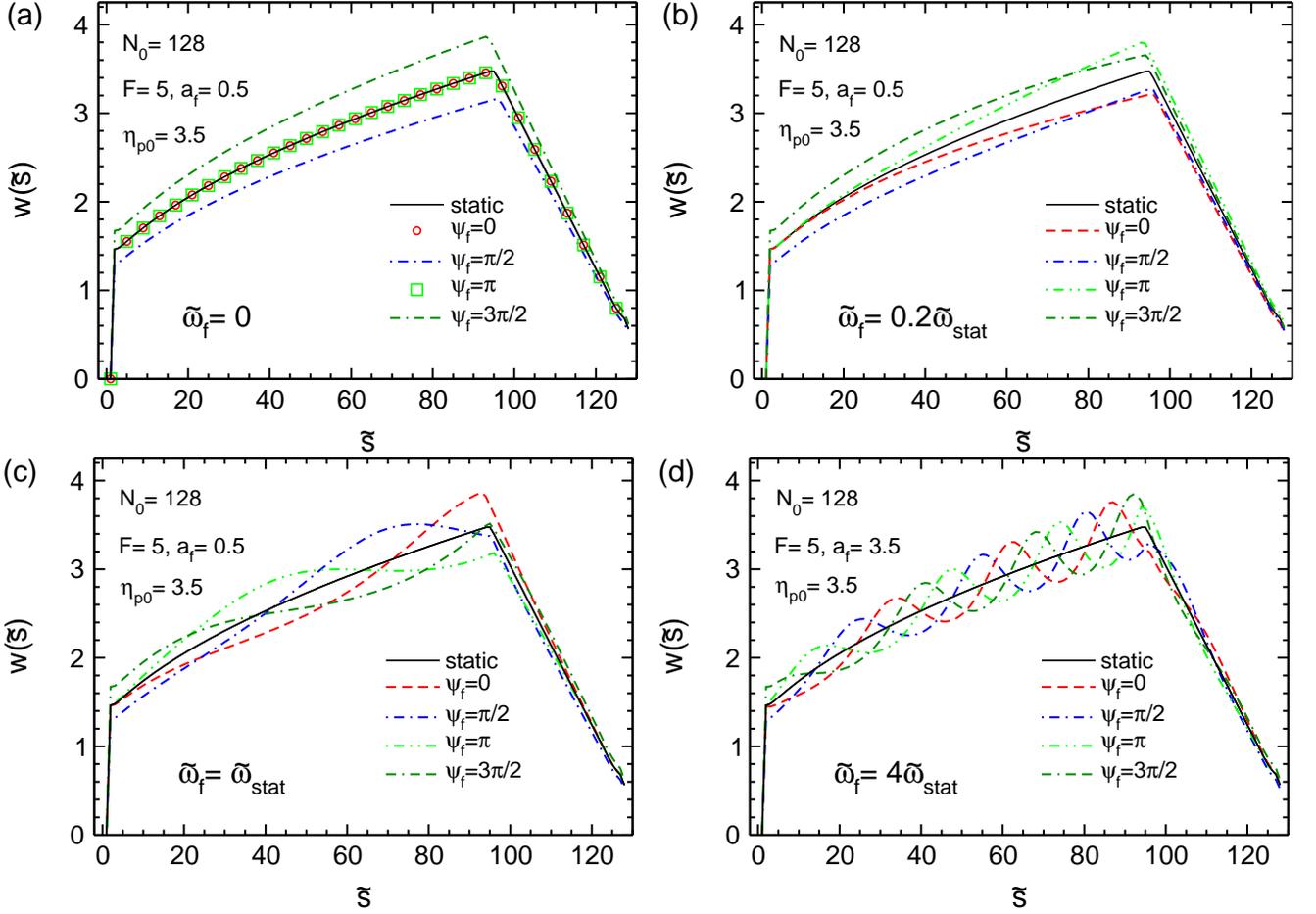
\begin{center}
    \begin{minipage}[b]{0.49\textwidth}\begin{center}
        \includegraphics[width=0.98\textwidth]{figure7a.eps}
    \end{center}\end{minipage} \hskip0.0cm
    \begin{minipage}[b]{0.49\textwidth}\begin{center}
        \includegraphics[width=0.98\textwidth]{figure7b.eps}
    \end{center}\end{minipage} \hskip0.0cm
    \begin{minipage}[b]{0.49\textwidth}\begin{center}
        \includegraphics[width=0.98\textwidth]{figure7c.eps}
    \end{center}\end{minipage} \hskip0.0cm
    \begin{minipage}[b]{0.49\textwidth}\begin{center}
        \includegraphics[width=0.98\textwidth]{figure7d.eps}
    \end{center}\end{minipage} \hskip0cm
\caption{(a) The waiting time, $w(\tilde{s})$, as a function of the translocation coordinate, $\tilde{s}$, 
for various values of initial force phases $\psi_f = 0, \pi/2, \pi$ and $3\pi/2$.
Here, the alternating external driving force is given by
$\tilde{f} (\tilde{t}) = \tilde{F} + \tilde{a}_{f} \sin (\tilde{\omega}_f \tilde{t} +\psi_f )$, where $F = 5$ and $a_f = 0.5$, 
the pore friction is constant, $\eta_{\textrm{p}0}= 3.5$, and the force frequency is $\tilde{\omega}_f = 0$.
Panels (b), (c) and (d) are the same as (a) but for different values of normalized
force frequencies $\tilde{\omega}_f / \tilde{\omega}_{\textrm{static}} = 0.2, 1$ and $4$, respectively. In all panels the chain length is 
$N_0 = 128$, and the black solid curve has been plotted for completely static case where the external driving 
force is constant, $F=5$, and the constant pore friction is $\eta_{\textrm{p}0}= 3.5$.} 
\label{fig:waiting_time}
\end{center}
\end{figure*}

To understand the effect of the oscillating pore and force on the translocation process, it is necessary to examine also the 
dynamics of the process and not just the resulting translocation time. Often the translocation process has been characterized 
by calculating the time evolution of the translocation coordinate $\tilde{s}$. However, a more sensitive measure is the monomer 
waiting time distribution that measures the time a monomer spends inside the pore. The tension propagation (increasing waiting 
time) and post propagation (decreasing waiting time) stages are also easy to distinguish from the distribution. For the oscillating 
force and static pore, the waiting time distributions are shown in Fig.~\!\ref{fig:waiting_time}. For the oscillating pore, 
the waiting times are not shown, but they are analogous to the oscillating force and consistent with the results discussed in 
the previous section.

In Fig.~\!\ref{fig:waiting_time}a the waiting time, $w(\tilde{s})$, is plotted as a function of the translocation coordinate, 
$\tilde{s}$, for various values of initial force phases $\psi_f = 0, \pi/2, \pi$ and $3\pi/2$. 
Here, the alternating external driving force is given by
$\tilde{f} (\tilde{t}) = \tilde{F} + \tilde{a}_{f} \sin (\tilde{\omega}_f \tilde{t} +\psi_f )$, where $F = 5$ and $a_f = 0.5$, 
the pore friction is constant, $\eta_{\textrm{p}0}= 3.5$, and the force frequency is $\omega_f = 0$.
The chain length is $N_0 = 128$, and the black solid curve has been plotted for the completely static case, where the external driving 
force is constant, $F=5$, and the constant pore friction is $\eta_{\textrm{p}0}= 3.5$.
With the initial phases $\psi_f = 0$ and $\pi$, the waiting time distribution curves collapse on the waiting time distribution 
curve for the static case, and as a result the corresponding translocation times for these two initial phases are exactly the 
same as it can be seen in Fig.~\!\ref{fig:translocation_time_alternating_force}. With $\psi_f = \pi/2$ the value of the force at 
$\tilde{\omega}_f = 0$ is greater than its value for the static case and therefore the waiting time is consistently below the static 
force curve and the overall translocation time is smaller. For $\psi_f = 3\pi/2$ the converse is true.

In Figs.~\!\ref{fig:waiting_time}b, c and d the waiting time distributions are shown for low, intermediate and high frequencies. 
In the low frequency regime, the waiting time gradually deviates from the static force distribution, while for the high frequency 
the distribution oscillates between the large and small force values. As the translocation time is the integral of the waiting time, 
it is clear from the high frequency distribution (Fig.~\!\ref{fig:waiting_time}d) why the effect of the oscillating force is 
averaged out. On the other hand, for the intermediate frequencies the waiting time and thus the translocation time can be higher 
or lower depending on the initial phase.

\begin{figure*}[t]
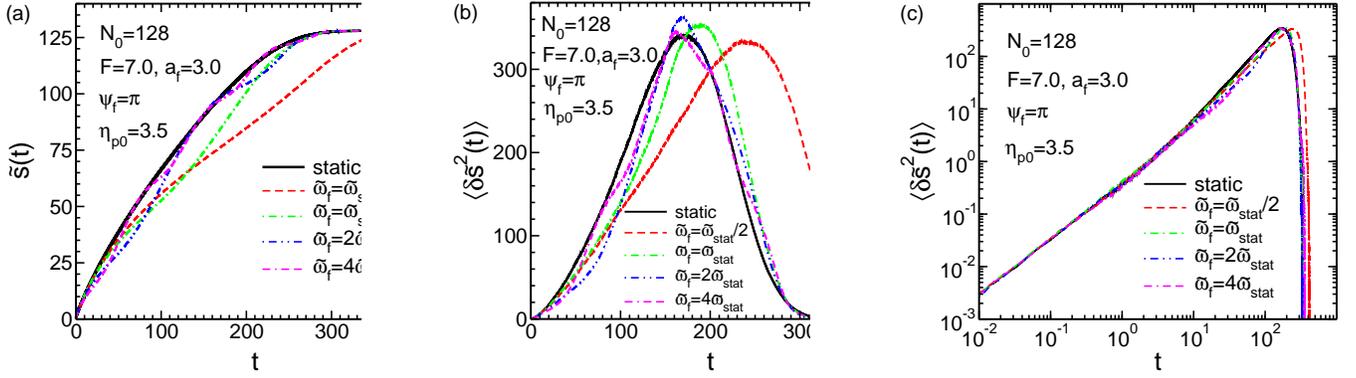
\begin{center}
    \begin{minipage}[b]{0.32\textwidth}\begin{center}
        \includegraphics[width=1.0\textwidth]{figure8a.eps}
    \end{center}\end{minipage} \hskip+0.1cm
    \begin{minipage}[b]{0.32\textwidth}\begin{center}
        \includegraphics[width=1.02\textwidth]{figure8b.eps}
    \end{center}\end{minipage} \hskip+0.1cm
    \begin{minipage}[b]{0.32\textwidth}\begin{center}
        \includegraphics[width=1.02\textwidth]{figure8c.eps}
    \end{center}\end{minipage} \hskip0cm
\caption{(a) The averaged translocation coordinate, $\tilde{s} (t)$, as a function of time, $t$, for the static 
case with constant external driving force $F=7$ and constant pore friction $\eta_{\textrm{p}0}=3.5$ (solid black curve), 
and also when the external driving force is alternating with 
$\tilde{f} (\tilde{t}) = \tilde{F} + \tilde{a}_f \sin ( \tilde{\omega}_f \tilde{t} + \psi_f )$ where $F=7$, $a_f = 3$ and
$\psi_f = \pi$, and the constant pore friction is $\eta_{\textrm{p}0} = 3.5$, for various values of force frequencies 
$\tilde{\omega}_f / \tilde{\omega}_{\textrm{stat}} = 0.5, 1, 2$ and $4$ (red dashed, green dashed-dotted, blue dashed-dotted-dotted 
and pink dashed-dashed-dotted curves, respectively). Panels (b) and (c) present the variance 
$\langle \delta \tilde{s}^2 ( t ) \rangle = \langle \tilde{s}^2 ( t ) \rangle - \langle \tilde{s} ( t ) \rangle ^ 2 $ 
as a function of time for the same parameters as of panel (a) in normal-normal and log-log 
scales, respectively. The chain length is $N_0 = 128$ in all panels.} 
\label{fig:time_evolution_of_s}
\end{center}
\end{figure*}

In addition to the waiting time, we have also solved for the time evolution of the translocation coordinate $\tilde{s}$. 
In this case, we have used the full stochastic model including the fluctuations due to thermal noise and initial configuration such that we 
can extract the variance of $\tilde{s}$ in addition to the mean value. In addition, we have also increased the driving force and the 
amplitude to make the oscillations more visible. The simulation parameters are thus $\tilde{f} (\tilde{t}) = \tilde{F} + 
\tilde{a}_f \sin ( \tilde{\omega}_f \tilde{t} + \psi_f )$, where $F=7$, $a_f = 3$,  $\psi_f = \pi$, pore friction $\eta_{\textrm{p}0} 
= 3.5$, and chain length $N_0 = 128$. The results for the mean value $\langle \tilde{s}(t) \rangle$ and the variance $\langle 
\delta \tilde{s}^2 ( t ) \rangle = \langle \tilde{s}^2 ( t ) \rangle - \langle \tilde{s} (t) \rangle^2$ 
are shown in Figs.~\!\ref{fig:time_evolution_of_s}a--c. The oscillations in the mean value of $\tilde{s}(t)$ are visible for 
the intermediate frequencies similar to the waiting time distribution. 

\section{Conclusions} \label{conclusions}

We have extended the recently developed iso-flux tension propagation theory to polymer translocation driven 
by a time-dependent external force and through an oscillating nanopore. Specifically, we take into account a 
periodically oscillating term in the external driving force and describe the flickering of the pore with a 
similar term in the pore friction. We have derived analytical approximations for the translocation time in 
the low and high frequency limits, and also in the intermediate frequency regime for small oscillation 
amplitudes. The analytical expressions can be used to interpret the numerical results obtained from solving 
the full model. The theory can also be easily generalized for other types of oscillations, including 
non-periodic ones.

We have shown that in the low and high frequency limits, the dynamics of the translocation process reduces to 
the non-oscillating cases; the translocation time is found to scale similarly to the static force and pore 
simulations, following the finite size scaling relation $\tilde{\tau} = c_1 N_0^{1+\nu} + c_2 \eta_{\textrm{p}} N_0$. 
However, in the intermediate frequency regime, determined by the characteristic time scale of the process, the effective 
scaling exponent, $\alpha (N_0) = d\ln \tilde{\tau} / d\ln N_0$, is shown to be extremely sensitive to small 
periodic fluctuations. The results indicate that such periodic fluctuations can indeed be used to either speed 
up or slow down the translocation process. Additionally, it is proposed that such control can be achieved either 
by tuning the oscillations in the driving force or in the pore characteristics. On the other hand, it is 
plausible that such oscillations may have significant influence on the experimentally measured translocation 
times and scaling exponents.

Our results shed light on the role that small oscillations in the external driving force and in the pore 
characteristics have on the dynamics of polymer translocation. The developed theory and the results may help 
in understanding both biological functions and experiments, which are being done on increasingly complex nanopore 
systems.

\begin{acknowledgments}
This work was supported by the Academy of Finland through its Centres
of Excellence Program (2012-2017) under Project No.~\!915804.
\end{acknowledgments}

\end{document}